\documentstyle[12pt]{article}
\def\appendix#1{
  \addtocounter{section}{1}
  \setcounter{equation}{0}
  \renewcommand{\thesection}{\Alph{section}}
 \section*{Appendix \thesection\protect\indent \parbox[t]{11.715cm} {#1}}
  \addcontentsline{toc}{section}{Appendix \thesection\ \ \ #1}
  }

\newcommand{\tr}[1]{\:{\rm tr}\,#1}
\def\const{{\rm const}}
\def\e{{\,\rm e}\,}
\def\Evac{E_{{\rm vac}}}
\def\evac{\epsilon_{{\rm vac}}}
\newcommand{\rf}[1]{(\ref{#1})}
\newcommand{\non}{\nonumber \\*}
\newcommand{\si}{\mathop{\rm si}\nolimits}
\newcommand{\ci}{\mathop{\rm ci}\nolimits}
\renewcommand{\l}[1]{\left.\frac{\partial }{\partial
\lambda }#1\right|_{\lambda =0}}
\begin{document}

\begin{titlepage}
\begin{flushright}
ITEP--TH--60/97\\
hep-th/9710235\\
October, 1997
\end{flushright}
\vspace{1.5cm}

\begin{center}
{\LARGE Ground State in Gluodynamics
\\[.4cm]
and Quark Confinement}\\
\vspace{1.9cm}
{\large K.~Zarembo}\\
\vspace{24pt}
{\it Institute of Theoretical and Experimental Physics,}
\\ {\it B. Cheremushkinskaya 25, 117259 Moscow, Russia} \\ \vskip .5 cm
E-mail: {\tt zarembo@itep.ru}
\end{center}
\vskip 2 cm
\begin{abstract}
The properties of the ground state wavefunctional in gluodynamics
responsible for confinement are considered. It is shown that confinement
arises due to the generation of a mass gap in the averaging over the
gauge group which is necessary to ensure the gauge invariance of the
ground state. The string tension is calculated in assumption of a
particular infrared behavior of the vacuum wavefunctional.
\end{abstract}

\end{titlepage}
\setcounter{page}{2}

\section{Introduction}

It is generally accepted that the confinement of quarks can be explained
as a consequence of the linear attractive potential arising between quark
and antiquark at large distances due to strong vacuum fluctuations of the
gluon fields. The origin of these strong fluctuations is a complicated
structure of the vacuum in QCD. In a literal sense, the vacuum is the
ground state of the Hamiltonian.  In this paper we consider the
concrete properties of the vacuum wavefunction in gluodynamics which
are responsible for confinement.

In principle, the ground state wavefunctional in QCD can be obtained by
solving Schr\"{o}\-din\-ger equation in some reasonable approximation.
The crucial point in any practical approach to this problem is the
maintenance of the gauge invariance.
The method of imposing gauge invariance condition by
integration over all gauges \cite{KK94, Cor88} appears to be most
convenient for our purposes. This method is familiar in the gauge
theories at finite temperature \cite{ft,Pol87}. It turns out that the
averaging over gauges provides a simple way to distinguish between
Coulomb and confining behavior of the quark-antiquark
potential. Confinement arises due to the absence of the long-range
correlations in the integration over the gauge group. The qualitative
explanation of the relationship between confinement and the generation
of a mass gap is given in Ref.~\cite{Pol87}.

\section{Ground state wavefunctional and gauge invariance}\label{2}

The canonical variables in pure gluodynamics (in the temporal gauge
$A_0=0$) are gauge potentials $A_i^A(x)$ and electric fields $E_i^A(x)$:
\begin{equation}\label{ccr}
[A_i^A(x),E_j^B(y)]=i\delta ^{AB}\delta _{ij}\delta (x-y).
\end{equation}
We consider gauge group $SU(N)$ and sometimes use matrix notations
for gauge potentials: $A_i=A_i^AT^A$, where $T^A$ are traceless
anti-Hermitean generators of $SU(N)$ normalized by $\tr T^AT^B=-\delta
^{AB}/2$.

The Hamiltonian of the Yang-Mills theory is
\begin{equation}\label{ham}
H=\int d^3x\,\left(\frac{1}{2}\,E_i^AE_i^A
+\frac{1}{4}\,F_{ij}^AF_{ij}^A\right),
\end{equation}
where $F_{ij}^A=\partial _iA_j^A-\partial _jA_i^A+gf^{ABC}A_i^BA_j^C$.
Apart from the Schr\"{o}dinger equation, the ground state wavefunction
satisfies also the Gauss law constraint:
\begin{equation}\label{gauss}
D_iE_i^A\, \Psi =0.
\end{equation}
The covariant derivative $D_i$ acts in the adjoint representation:
$D_i^{AB}=\delta ^{AB}\partial _i+gf^{ACB}A_i^C$.

In the ``coordinate'' representation, the electric field operators act as
the variational derivatives, $E_i^A(x)=-i\,\delta /\delta A_i^A(x)$, and
the wavefunction $\Psi $ is a functional of the gauge potentials.
The operator on the left hand side of Eq.~\rf{gauss} generates gauge
transformations
\begin{equation}\label{AU}
A_i\longrightarrow A_i^\Omega =
\Omega ^{\dagger}\left(A_i+\frac{1}{g}\,\partial
_i\right)\Omega ,
\end{equation}
so the Gauss law \rf{gauss} is the
infinitesimal form of the gauge invariance condition:
\begin{equation}\label{ginv}
\Psi [A^\Omega ]=\Psi [A].
\end{equation}
Strictly speaking, this is true only for topologically trivial gauge
transformations -- the wavefunctional of the $\theta $-vacuum is
invariant only up to a phase factor $\e^{ik\theta}$,
where $k$ is a winding number of the gauge transformation, but for
simplicity we assume that $\theta =0$, then Eq.~\rf{ginv} holds for any
$\Omega (x)$.

In principle, any functional of $A_i$ can be made gauge invariant by
projection on the subspace of states obeying the Gauss law. The
projection can be realized by averaging over the gauge group
\cite{ft}. So, the following ansatz for the ground state wavefunctional
automatically  satisfies \rf{ginv}:
\begin{equation}\label{ansatz}
\Psi [A]=\int [DU]\,\e^{-S[A^U]},
\end{equation}
The invariance of this expression follows simply from the fact that a
gauge transformation $A_i\rightarrow A_i^\Omega $ can be absorbed by the
change of integration variables: $U\rightarrow \Omega ^{\dagger}U$.

This method of imposing gauge invariance was used in the variational
approach of Ref.~\cite{KK94}. The variational ansatz of Ref.~\cite{KK94}
consists of restricting to the purely quadratic action
\begin{equation}\label{defk}
S_0[A]=\frac{1}{2}\int d^3xd^3y\,A_i^A(x)K(x-y)A_i^A(y).
\end{equation}
with some special choice of the coefficient function $K(x-y)$.
We consider more general situation when the action $S[A]$ can contain
arbitrary powers of gauge potentials.  However, it is clear that the
functional $S[A]$ is defined ambiguously.  Different choices of $S$ may
lead to one and the same $\Psi$ and it is convenient to get rid of this
arbitrariness demanding the quadratic part of $S[A]$ to coincide with
$S_0[A]$. The color and the Lorentz structure of the quadratic form in
\rf{defk} is dictated by rotational, translational and global gauge
symmetries. The general principles allow also a longitudinal term, but it
can be always removed without change of the wavefunctional, as shown in
Appendix.

The introduction of the ``bare'' wavefunctional
$\exp(-S[A])$, which is not gauge invariant and requires the
averaging over gauges, nevertheless, has several advantages.
There is a hope that the bare wavefunctional looks rather simple in some
reasonable approximation \cite{KK94}, while after the integration over
the gauge group one obtains sufficiently complicated gauge invariant
functional.  The possibility to impose the gauge invariance condition
exactly on an approximate wavefunctional of a simple form makes this
method convenient for variational calculations \cite{KK94}.  We shall
show that the construction of states containing static color charges is
straightforward in this framework, which allows to formulate a relatively
simple criterion of confinement.

Due to the gauge invariance the normalization integral
$\left\langle\Psi |\Psi \right\rangle$ contains a group volume factor.
This factor is easily extracted in the representation \rf{ansatz}:
\begin{equation}\label{norm}
\left\langle\Psi |\Psi \right\rangle=\int [DU][DU'][dA]\,
\e^{-S[A^U]-S[A^{U'}]}=\int [DU][dA]\,\e^{-S[A^U]-S[A]},
\end{equation}
where $U'$ is removed by the change of integration variables:
$A_i\rightarrow A_i^{U'^{\dagger}}$, $U\rightarrow U'U$. The same trick
works for the expectation value of any gauge invariant operator
\cite{KK94}.  In particular,
\begin{equation}\label{meanh}
\left\langle\Psi |H|\Psi \right\rangle= \int [DU][dA]\,\int
d^3x\,\left(\frac{1}{2}\,\frac{\delta ^2S[A]}{\delta A_i^A\delta
A_i^A}-\frac{1}{2}\,\frac{\delta S[A]}{\delta A_i^A}
\,\frac{\delta S[A]}{\delta A_i^A}
+\frac{1}{4}\,F_{ij}^AF_{ij}^A\right)\e^{-S[A^U]-S[A]}.
\end{equation}
Thus, the vacuum energy is determined by the average of the operator
\begin{equation}\label{defr}
R[A]=\int d^3x\,\left(\frac{1}{2}\,\frac{\delta
^2S[A]}{\delta A_i^A\delta
A_i^A}-\frac{1}{2}\,\frac{\delta S[A]}{\delta A_i^A}
\,\frac{\delta S[A]}{\delta A_i^A}
+\frac{1}{4}\,F_{ij}^AF_{ij}^A\right)
\end{equation}
in the statistical ensemble which is defined by
the partition function \rf{norm}.  The same result is obtained by
averaging of $R[A^U]$ -- it arises when the variational derivatives in
the Hamiltonian act not on the right but on the left. It is
natural to preserve the symmetry of \rf{norm} under the interchange of
$A_i$ with $A_i^U$ and to average the symmetric combination
$\frac{1}{2}\Bigl(R[A^U]+R[A]\Bigr)$. Since the vacuum energy is
proportional to the volume, it is also natural to add this operator to
the action and to consider the partition function
\begin{equation}\label{defz}
Z=\int
[DU][dA]\,\e^{-S[A^U]-S[A]+\frac{\lambda }{2}\,R[A^U]+\frac{\lambda
}{2}\,R[A]}.
\end{equation}
The physical parameters characterizing the QCD vacuum are simply related
to the thermodynamic quantities in the statistical system defined by this
partition function. At least, this is true for the gluon condensate and
the string tension, as shown below.

The vacuum energy in QCD is equal to the derivative of the free energy
$\ln Z$ with respect to $\lambda $:
\begin{equation}\label{evac}
\Evac=\frac{\left\langle\Psi |H|\Psi \right\rangle}
{\left\langle\Psi |\Psi \right\rangle}=\l{\ln Z}.
\end{equation}
It is worth mentioning that the perturbation in \rf{defz} softens the
short-distance behavior, since $R$ is constructed from dimension four
operators.  So, the parameter $\lambda $, which has the dimension of
length, can be regarded also as an UV cutoff.

The regularized energy density, $\evac=\Evac/V-\mbox{UV divergent terms}$,
can be expressed through the gluon condensate \cite{SVZ79}:
\begin{equation}\label{edens}
\evac=\frac{\beta (\alpha _s)}{4\alpha _s}\left\langle
0\left|F_{\mu \nu }^AF^{A\mu \nu}\right|0\right\rangle,
\end{equation}
where $\alpha _s=g^2/4\pi$ and $\beta (\alpha _s)=-11N\alpha
_s^2/6\pi+\ldots$. The gluon condensate is known to be positive. Hence,
the regularized
vacuum energy density is negative. Therefore, we expect that $R$ is
negative definite operator after the subtraction of a field
independent divergent constant $\frac{1}{2}VK(0)=V\int \frac{d^3p}{(2\pi
)^3}\,\frac{K(p)}{2}$ -- zero point energy which comes from the second
variation of the quadratic term in $S[A]$. This explains why
the sign before $\lambda $ in Eq.~\rf{defz} is positive.

\section{Charged states and confinement}

In the presence of charges, the Gauss law \rf{gauss} acquires a right
hand side proportional to the charge density.
We consider the case of two static charged sources placed
at the points $x_1$ and $x_2$ with quark and antiquark quantum numbers.
The wavefunctional of such state transforms in $N\otimes\bar{N}$
representation of $SU(N)$.  The Gauss law constraint for it is
\begin{equation}\label{gauss'}
D_iE_i^A(x)\, \Psi_{ab}=ig\Bigl(\delta (x-x_2)T^A_{b'b}\delta _{aa'}
-\delta (x-x_1)T^A_{aa'}\delta _{b'b}\Bigr)\Psi _{a'b'},
\end{equation}
which implies the following transformation law:
\begin{equation}\label{ginv'}
\Psi_{ab} [A^\Omega;x_1,x_2]=\Omega ^{\dagger}_{aa'}(x_1)\Omega
_{b'b}(x_2)\Psi_{a'b'} [A;x_1,x_2].
\end{equation}

The construction of the state $\Psi _{ab}[A;x_1,x_2]$ requires
operators in the fundamental representation. These operators are
necessarily nonlocal in terms of the gauge fields, since gluons transform
in the adjoint representation. The integral formula \rf{ansatz} for
the wavefunctional, in fact, introduces the fields in the
fundamental representation, since $U$ transforms as $U\rightarrow\Omega
^{\dagger}U$.  This circumstance allows to construct the states with
quark quantum numbers entirely from local operators at a price of
the subsequent integration over all gauges.  The state obeying \rf{ginv'}
is constructed as follows:
\begin{equation}\label{ansatz'}
\Psi_{ab}[A;x_1,x_2]=\int [DU]\,U_{ac}(x_1)U^{\dagger}_{cb}(x_2)
\e^{-S[A^U]}.
\end{equation}
The crucial point here is that the action $S$ in bulk is the same as
in the vacuum sector. The reasoning is based on the variational
arguments.

The energy of the state \rf{ansatz'} consists of the vacuum energy and
quark-antiquark interaction potential:
\begin{equation}\label{e12}
\frac{\left\langle\Psi_{ab}|H|\Psi_{ab}\right\rangle}
{\left\langle\Psi_{ab} |\Psi_{ab} \right\rangle}
=\Evac+V(x_1-x_2).
\end{equation}
The vacuum energy is proportional to the volume, while the interaction
potential remains finite in the infinite volume limit. Suppose we are
trying to minimize the energies of the states $\Psi $ and $\Psi _{ab}$.
Then $O(V)$ terms in the variational equations are the same in both
cases.  Since variational equations completely determine the ground state
in each sector, the action in Eq.~\rf{ansatz'} coincides with that in
Eq.~\rf{ansatz} in bulk. In principle, in the charged sector, some extra
terms localized near the points $x_1$ and $x_2$ can arise.  It is more
natural to include these terms in the definition of the operators
$U_{ac}(x_1)$ and $U^{\dagger}_{cb}(x_2)$, which create external charges.
So, strictly speaking, for the state \rf{ansatz'} to be an eigenfunction
of the Hamiltonian, the operators $U_{ac}(x_1)$ and
$U^{\dagger}_{cb}(x_2)$ should be replaced by some dressed ones ${\cal
U}_{ac}(x_1)$ and ${\cal U}^{\dagger}_{cb}(x_2)$ with the same quantum
numbers. The particular form of these dressed operators is determined by
$O(1)$ terms in the variational equations. However, in what follows we do
not distinguish between $U$ and ${\cal U}$, since we are interested in a
large distance behavior, which is determined by the
bulk theory and thus is independent of the explicit form of the operators.

The energy of the state \rf{ansatz'} can be expressed in terms
of correlation functions in the statistical system defined by
the partition function \rf{defz}. Analogously to Eqs.~\rf{norm},
\rf{meanh} we find:
\begin{equation}\label{norm'}
\left\langle\Psi_{ab} |\Psi_{ab} \right\rangle
=\int [DU][dA]\,\tr U(x_1)U^{\dagger}(x_2)\e^{-S[A^U]-S[A]},
\end{equation}
\begin{equation}\label{meanh'}
\left\langle\Psi_{ab} |H|\Psi_{ab} \right\rangle= \int [DU][dA]\,
\tr U(x_1)U^{\dagger}(x_2)
\,\frac{1}{2}\Bigl(R[A^U]+R[A]\Bigr)\e^{-S[A^U]-S[A]}.
\end{equation}
These expressions together with Eq.~\rf{evac} lead to the following
representation for the $q\bar{q}$ potential:
\begin{equation}\label{v12}
V(x_1-x_2)=\l{\ln\left\langle\tr U(x_1)U^{\dagger}(x_2)\right\rangle},
\end{equation}
where the average is defined as usual:
\begin{equation}\label{aver}
\left\langle{\cal O}\right\rangle
=\frac{1}{Z}\int [DU][dA]\,{\cal O}
\e^{-\Gamma [A,U]},
\end{equation}
\begin{equation}\label{defg}
\Gamma [A,U]=S[A^U]+S[A]-\frac{\lambda }{2}\,R[A^U]-\frac{\lambda
}{2}\,R[A].
\end{equation}
All subsequent consideration is based on this result.

The averaging over the gauge group in \rf{aver}  may or may not produce a
mass gap. Consider first the former case. Then the two-point correlator in
Eq.~\rf{v12}  falls exponentially at  large distances:
\begin{equation}\label{expfall}
\left\langle\tr U(x_1)U^{\dagger}(x_2)\right\rangle
=\frac{C}{r^\eta}\,\e^{-mr}+\dots,
\end{equation}
where $r=|x_1-x_2|$, $m$ is a mass gap and $\eta$ is related to an
anomalous dimension of the operator $U$.  Note that the constant term  in
the large distance asymptotics \rf{expfall} is  forbidden by the symmetry
$U(x)\rightarrow U(x)h$, which can not be broken spontaneously, since
this symmetry follows from the global color invariance of the action
$S[A]$.

Substituting \rf{expfall} in Eq.~\rf{v12}
we find that $q\bar{q}$ potential grows linearly at large distances:
\begin{equation}\label{qq}
V(r)=\sigma r+\ldots
\end{equation}
with
\begin{equation}\label{strten}
\sigma =-\left.\frac{\partial m}{\partial \lambda}\right|_{\lambda =0}.
\end{equation}
Thus, the generation of a mass gap in the averaging over gauges leads
to the confinement of charges in the fundamental representation. It
is worth mentioning that the mass gap, and thus the string tension
$\sigma $, is determined by the bulk theory. This means that the
confinement actually is the property of the ground state and not of the
operators creating external charged sources.

The screening length in Eq.~\rf{expfall} depends only on the quantum
numbers of the operators on the left hand side. This justifies the use
of the bare operators $U$ instead of the dressed ones ${\cal U}$ in
the study of the large distance behavior of the $q\bar{q}$ potential. The
above arguments are essentially based on the absence of local operators
with quark quantum numbers in the pure gluodynamics and they fail for the
states with adjoint charges. In this case, the state \rf{ansatz'} (with
the matrices $U$ and $U^{\dagger}$ taken in the adjoint representation)
inevitably mixes with the ones of the form ${\cal O}^A(x_1){\cal
O}^B(x_2)\Psi $, where ${\cal O}^A$ are some local operators, for
example, ${\cal O}^A=F_{ij}^A$. The energy of such states, roughly
speaking, is determined by Eq.~\rf{v12} with
$\left\langle\tr U(x_1)U^{\dagger}(x_2)\right\rangle$ replaced by
$\left\langle{\cal O}^2(x_1){\cal O}^2(x_2)\right\rangle$.  The large
distance asymptotics of this correlator is determined by an expectation
value of ${\cal O}^2$, which is generally nonzero. It is easy to see that
appearance of the constant term leads to the screening of the
interaction potential:  $V_{{\rm adj}}(r)\sim\e^{-mr}$.  However, the
mixing with the states composed from purely gluonic operators can be
small, then the potential for adjoint charges will exhibit a linear
growth at moderate distances.

Another remark concerns the anomalous dimension
$\eta $ entering
Eq.~\rf{expfall}. It gives rise to the logarithmic term in the $q\bar{q}$
potential, unless
\begin{equation}\label{adim}
\left.\frac{\partial \eta }{\partial \lambda }\right|_{\lambda =0}=0.
\end{equation}
It is generally assumed that subleading terms in the $q\bar{q}$ potential
decrease at infinity, so one may expect
that Eq.~\rf{adim} holds for the ground state in gluodynamics.

Now we turn to the case when the mass gap is absent -- $m=0$ in
Eq.~\rf{expfall}. Assuming the validity of \rf{adim} we find that the
leading term in the large distance asymptotics of the pair correlator
is insufficient to determine the behavior of the $q\bar{q}$
interaction potential. Retaining the next term:
\begin{equation}\label{nlo}
\left\langle\tr U(x_1)U^{\dagger}(x_2)\right\rangle
=\frac{C}{r^\eta}\left(1-\frac{B}{r}+\ldots\right),
\end{equation}
we obtain the Coulomb law:
\begin{equation}\label{coul}
V(r)=\const-\frac{Q^2}{4\pi r}+\ldots,~~~~Q^2=4\pi \left.\frac{\partial
B}{\partial \lambda }\right|_{\lambda =0}.
\end{equation}
Therefore, the presence of long-range correlations in the averaging over
gauges leads to the Coulomb interaction of static charges. Unfortunately,
there is no symmetry that distinguish between Coulomb and confinement
phases, in contrast to the finite temperature gauge theories
\cite{ft,Pol87}.

\section{Examples}

Below the general formalism of the previous section is illustrated by two
examples.  One is $U(1)$ gauge theory. Another refers directly to
gluodynamics. We show that minor assumptions about an infrared behavior
of the wavefunctional allow to calculate the string tension.

\subsection{QED}

In the Abelian theory $U=\e^{ig\varphi }$, $A^U_i=A_i+\partial _i\varphi
$ and the action $S[A]$ is quadratic, i.e.\ coincides with \rf{defk}, and
the averaging over gauges reduces
to Gaussian integrals. In the vacuum sector, the integration over
$\varphi $ in Eq.~\rf{ansatz} replaces $A_i$ by $A_i^\bot=\left(\delta
_{ij}-\frac{\partial _i\partial _j}{\partial ^2}\right)A_j$:
\begin{equation}\label{qed} \Psi
[A]=\exp\left(-\frac{1}{2}\int d^3xd^3y\,A_i^\bot(x)K(x-y)A_i^\bot(y)
\right).
\end{equation}
This is an exact ground state wavefunctional of
the free electromagnetic
field, provided that $K=\sqrt{-\partial ^2}$ (in the
momentum space $K(p)=|p|$). The integration over the gauge group for the
state \rf{ansatz'} containing charged sources also can be easily
performed yielding:
\begin{equation}\label{dstate} \Psi
[A;x_1,x_2]=\e^{igV(x_1)}\e^{-igV(x_2)}\Psi[A], \end{equation} where
\begin{equation}\label{defV}
V(x)=\int d^3y\,\frac{1}{-\partial ^2}(x-y)\partial _iA_i(y).
\end{equation}
The operators $\e^{iqV(x)}$ were introduced by Dirac \cite{Dir55}. The
states of type \rf{dstate} created by Dirac charge operators behave
properly  under gauge transformations and are the eigenstates of the
Hamiltonian of the free electromagnetic field. The wavefunctional
\rf{ansatz'} provides a natural nonabelian generalization of
this construction.

Dropping an irrelevant constant term we get for the
action \rf{defg} in the Abelian theory:
\begin{equation}\label{gab}
\Gamma=\frac{1}{2}\int d^3xd^3y\,A_i(x)
\left(K(x-y)+\frac{\lambda }{2}\,K^2(x-y)\right)A_i(y)
+(A_i\longleftrightarrow A_i+\partial _i\varphi )-
\frac{\lambda }{4}\int d^3x\,F_{ij}F_{ij}.
\end{equation}
Here $K^2(x-y)$ denotes the kernel of the operator $K^2$.
The gauge potentials can be integrated out by the change
of variables:
\begin{equation}\label{shift}
A_i=\bar{A}_i-\frac{1}{2}\,\partial _i\varphi .
\end{equation}
The shifted gauge potentials $\bar{A}_i$ decouple and we are left with
the effective action for $\varphi $:  \begin{equation}\label{seff}
\Gamma_{\rm eff}
=\frac{1}{4}\int d^3xd^3y\,\partial _i\varphi(x)
\left(K(x-y)+\frac{\lambda }{2}\,K^2(x-y)\right)\partial _i\varphi(y).
\end{equation}
So, for the two-point correlator we get:
\begin{equation}\label{uu+}
\left\langle\e^{ig\varphi (x_1)}\e^{-ig\varphi (x_2)}\right\rangle
=\e^{g^2(D(x_1-x_2)-D(0))},
\end{equation}
where $D$ is the propagator of the field $\varphi $:
\begin{equation}\label{d-1}
D^{-1}=\frac{-\partial ^2}{2}\left(K+\frac{\lambda }{2}\,K^2\right).
\end{equation}
Differentiating the propagator with respect to $\lambda $ we find:
\begin{equation}\label{dddl}
\left.\frac{\partial D}{\partial \lambda }\right|_{\lambda =0}=
-\frac{1}{-\partial ^2}.
\end{equation}
Hence, according to Eq.~\rf{v12}, we obtain the Coulomb law for the
interaction potential:
\begin{equation}\label{1/r}
V(x_1-x_2)=-g^2\left(\frac{1}{-\partial ^2}(x_1-x_2)
-\frac{1}{-\partial ^2}(0)\right)=-\frac{g^2}{4\pi r}
+\mbox {self-energy}.
\end{equation}
It is interesting that the explicit form of the coefficient
function $K$ was not used anywhere in the derivation of this result.
Nevertheless, it is useful to substitute $K=\sqrt{-\partial ^2}$ in
Eq.~\rf{d-1} in order to compare \rf{uu+} with \rf{nlo}. After some
calculations we obtain:
\begin{eqnarray}\label{abex}
\left\langle\e^{ig\varphi (x_1)}\e^{-ig\varphi (x_2)}\right\rangle
&=&\left(\frac{\lambda \e^{-\gamma }}{2r}\right)^{\frac{g^2}{\pi^2 }}
\e^{-\frac{\lambda g^2}{2\pi ^2r}\left(\frac{\pi
}{2}+\cos\frac{2r}{\lambda }\,\si\frac{2r}{\lambda }
-\sin\frac{2r}{\lambda }\,\ci\frac{2r}{\lambda }\right)}
\non
&=&\left(\frac{\lambda \e^{-\gamma }}{2r}\right)^{\frac{g^2}{\pi^2 }}
\left(1-\frac{\lambda g^2}{4\pi r}+\ldots\right),
\end{eqnarray}
where $\gamma $ is the Euler constant. Comparing this expression with
\rf{nlo} we find that the constant $C$ diverges in the limit $\lambda
\rightarrow 0$. This divergence reflects the fact that the self-energy of
the Coulomb charge is infinite, but this infinity is regularized by
$\lambda $.

\subsection{QCD}

Due to the asymptotic freedom, the short distance behavior of the
$q\bar{q}$ potential is determined by the perturbation theory. At weak
coupling, one can expand $U=\e^{g\Phi ^AT^A}=1+g\Phi ^AT^A+\ldots$. Only
the quadratic term in the action $S[A]$ contribute to the leading order
in $g$ and the calculations are literally the same as in the Abelian
theory. Since the explicit form of the coefficient function $K$ is
inessential for these calculations, at short distances we get the
Coulomb potential \rf{1/r} with $g^2$ multiplied by a group factor
$(N^2-1)/N$.

It is naturally to assume that at large distances
the long wavelength approximation is
valid and the action \rf{defg} can be approximated by its derivative
expansion.  The structure of the derivative expansion for $\Gamma $
depends on the behavior of the coefficient function $K(p)$ at small
momenta. Again, only the quadratic term in the action is essential, since
the remaining ones correspond to operators of higher dimension.
Here we assume that $K(p=0)\neq 0$, so
\begin{equation}\label{k(0)}
K(p)=M+O(p^2).
\end{equation}
Neglecting $O(p^2)$ terms we obtain:
\begin{equation}\label{gamma}
\Gamma=\frac{\mu }{2} \int d^3x\,\left(A_i^AA_i^A
+A_i^{U\,A}A_i^{U\,A}\right),
\end{equation}
where
\begin{equation}\label{mu}
\mu =M+\frac{\lambda M^2}{2}.
\end{equation}
The magnetic term $(F_{ij}^A)^2$ is also irrelevant in this
approximation, since it contains extra derivatives in comparison to
$(A_i^A)^2$.

The case of purely quadratic action $S$ with the coefficient
function $K$ approximated by a constant at small momenta was considered
in Ref.~\cite{KK94}. Following this consideration, which is based on
methods typical for nonlinear sigma-models, we
express the string tension in terms of $M$ and the gauge
coupling constant.

First, we get rid of the integration over $A_i$ by the change of
variables:
\begin{equation}\label{shift'}
A_i=\bar{A}_i-\frac{1}{2g}\,\partial _iUU^{\dagger},
\end{equation}
which is a counterpart of \rf{shift}. After this, $\bar{A}_i$ decouples
and we get the effective action for $U$:
\begin{equation}\label{geff}
\Gamma _{\rm eff}=-\frac{\mu }{2g^2}\int
d^3x\,\tr\left(U^{\dagger}\partial _iU\right)^2
=\frac{\mu }{2g^2}\int d^3x\,\tr\partial _iU^{\dagger}\partial _iU.
\end{equation}
It is convenient to integrate over $U(N)$ instead of $SU(N)$. This can be
achieved by adding the following term to the action:
\begin{equation}\label{g1}
\Gamma _1=\frac{\mu }{2g^2N}\int d^3x\,\left(\tr
U^{\dagger}\partial _iU\right)^2.
\end{equation}
Since $\Gamma _{\rm eff}+\Gamma _1$ is invariant under phase
transformations $U\rightarrow\e^{i\varphi }U$, it depends only on $SU(N)$
variables, for which it reduces to $\Gamma _{\rm eff}$. The last step
consists in the replacement of the group integral  by an integral over
complex $N\times N$ matrices. The unitarity condition is imposed by
means of a Lagrange multiplier:
\begin{equation}\label{lagrange}
\Gamma _2=\frac{\mu }{2g^2}\int d^3x\,\tr\sigma
\left(U^{\dagger}U-1\right),
\end{equation}
where
\begin{equation}\label{sigma}
\sigma _{ab}=m^2\delta _{ab}+v_{ab},
\end{equation}
and $v$ is anti-Hermitean matrix.

The crucial point is that \rf{lagrange} introduces a mass term for $U$,
so the pair correlator falls exponentially at large
distances as  in Eq.~\rf{expfall}, which leads to confinement. The mass
gap $m$ is determined by the equation of motion for $\sigma $. This
equation follows from the effective action obtained by integrating out
$U(x)$.  The gap equation at the same time is equivalent to the unitarity
condition $\left\langle U(x)U^{\dagger}(x)\right\rangle=1$. In the
one-loop approximation, we have
\begin{equation}\label{gap}
\int\frac{d^3p}{(2\pi)^3}\,\frac{1}{p^2+m^2}=
\frac{\mu}{2g^2N}.
\end{equation}
The left hand side is UV divergent and, in principle, it is necessary to
introduce a cutoff in order to find the mass gap \cite{KK94}.
However, the string tension appears to be cutoff independent, since
for the derivative of $m$ we obtain:
\begin{equation}\label{dgap}
-\frac{\partial m^2}{\partial \lambda}
\int\frac{d^3p}{(2\pi)^3}\,\frac{1}{(p^2+m^2)^2}=\frac{M^2}{4g^2N},
\end{equation}
and, according to Eq.~\rf{strten}, the string tension is
\begin{equation}\label{strten'}
\sigma=-\frac{\partial m}{\partial \lambda}=\frac{\pi M^2}{g^2N}
=\frac{M^2}{4\alpha_sN}.
\end{equation}

This result is not universal, it is valid under the condition that
$K(p=0)\neq 0$. This assumption is compatible with the variational
estimates of Ref.~\cite{KK94}. However, the arguments in favor of
$K(p=0)=0$ also exist, and there are some indications \cite{Gre79} that
the expansion of $K(p)$ at small momenta begins with the term of order
$p^2$.  If such behavior actually holds, the present treatment of
confinement properties should be modified.

\section{Discussion}

Our main result is that, in the Hamiltonian picture, the confinement
arises as a consequence of the generation of a mass gap in the averaging
over gauges.  Although it is possible to derive the linear potential and
even to calculate the string tension under some simple and not very
restrictive assumptions about the infrared properties of the ground state
in gluodynamics, the quantitative treatment requires a more detailed
knowledge of the vacuum wavefunctional. Possibly, the variational
techniques \cite{KK94} can provide a reasonable approximation for the
ground state in QCD incorporating both asymptotic freedom \cite{BK97} and
confinement.

A more detailed quantitative treatment of the ground state in
gluodynamics may provide insights in some other nonperturbative
properties of the low-energy QCD. In particular, the usual picture of the
$q\bar{q}$ interaction is associated with a string
 stretched between quark and antiquark.  Perhaps, this
picture would correspond to a sum-over-path representation for the
correlator entering Eq.~\rf{v12}:  \begin{equation}\label{path}
\left\langle\tr U(x_1)U^{\dagger}(x_2)\right\rangle
=\int_{X(0)=x_1}^{X(1)=x_2} [dX]\,\e^{-{\cal H}[X(\xi )]}.
\end{equation} The large distance asymptotics \rf{expfall} is naturally
interpreted as a saddle point approximation for the path integral
\rf{path} and the functional ${\cal H}[X(\xi )]$ has a meaning of an
effective string Hamiltonian.

\subsection*{Acknowledgments}

This work was supported in part by
 CRDF grant \mbox{96-RP1-253},
 INTAS grant \mbox{96-0524},
 RFFI grant \mbox{97-02-17927}
 and grant \mbox{96-15-96455} of the support of scientific schools.

\setcounter{section}{0}
\setcounter{subsection}{0}
\appendix{Removing arbitrariness in the action}

To demonstrate that the representation \rf{ansatz} is ambiguous let us
consider a set of functionals $S[A;\xi ]$, whose dependence on the
parameter $\xi $ is determined by the equation
\begin{equation}\label{hc}
\frac{\partial S}{\partial \xi }=\int d^3xd^3y\,A_i^A(x)W(x-y)\partial
_iD_j^{AB}\frac{\delta S}{\delta A_j^B(y)}.
\end{equation}
Here $W(x-y)$ is an arbitrary function.
Then the wavefunctional
\begin{equation}\label{ans'}
\Psi [A;\xi ]=\int [DU]\,\e^{-S[A^U;\xi ]}
\end{equation}
is independent of $\xi $ up to a normalization factor, which is
inessential.  The proof is based on Schwinger-Dyson equations.

The measure of integration in \rf{ans'} is invariant under left shifts:
\begin{equation}\label{left}
\delta _\omega U=U\omega ,~~~~\delta _\omega U^{\dagger}=-\omega
U^{\dagger},
\end{equation}
where $\omega $ is traceless anti-Hermitean matrix. The
twisted gauge potentials
$A_i^U$ transform under left shifts as
\begin{equation}\label{left'}
\delta _\omega A_i^U=[A_i^U,\omega ]+\frac{1}{g}\,\partial
_i\omega\equiv\frac{1}{g}\,D_i^U\omega .
\end{equation}
The Schwinger-Dyson equation follows from the identity
\begin{eqnarray}\label{schd}
0&=&\int [DU]\,\delta _\omega \left(\partial _iA_i^{U\,A}(x)
\e^{-S[A^U]}\right) \non
&=&\frac{1}{g}\int [DU]\,\left(\partial _iD_i^{U\,AC}\omega ^C(x)
+\partial _iA_i^{U\,A}(x)\int d^3y\,\omega ^C(y)D_j^{U\,CB}
\frac{\delta S[A^U]}{\delta A_j^{U\,B}(y)}\right)\e^{-S[A^U]}.
\nonumber
\end{eqnarray}
This equality holds for any $\omega^C(y)$. Taking
$\omega ^C(y)=\delta ^{CA}W(x-y)$ and summing over ${\scriptstyle A}$
we find:
\begin{equation}\label{maineq}
(N^2-1)\partial ^2W(0)\,\Psi [A]+\int [DU]\,\int d^3y\,\partial
_iA_i^{U\,A}(x)W(x-y)D_j^{U\,AB}
\frac{\delta S[A^U]}{\delta A_j^{U\,B}(y)}\,\e^{-S[A^U]}=0.
\end{equation}
Combining this equation with \rf{hc} we obtain:
\begin{equation}\label{he}
\frac{\partial \Psi }{\partial \xi }=-\int [DU]\,\int
d^3xd^3y\,A_i^{U\,A}(x)W(x-y)\partial _iD_j^{U\,AB}\frac{\delta S}{\delta
A_j^{U\,B}(y)}\,\e^{-S[A^U]}=c\Psi ,
\end{equation}
where $c=(N^2-1)V\partial ^2W(0)$. Thus,
\begin{equation}\label{psixi}
\Psi [A;\xi ]=\e^{\xi c}\Psi [A;0].
\end{equation}

The above arguments show that the solution of Eq.~\rf{hc} for any $W$ and
at arbitrary value of the parameter $\xi $ gives essentially the same
wavefunctional. The action $S$ can be expanded in the power series:
\begin{equation}\label{te}
S[A]=\frac{1}{2}\int d^3xd^3y\,A_i^A(x)K_{ij}(x-y)A_j^A(y)+\ldots\,.
\end{equation}
Then the equality \rf{hc} is rewritten as an infinite set of equations for
coefficient functions. The first equation in this set contains only the
two-point function and reads:
\begin{equation}\label{hck}
\frac{\partial K_{ij}}{\partial \xi }=W\partial _i\partial _kK_{kj}.
\end{equation}
Here $W$ denotes an integral operator with the kernel $W(x-y)$. We can
start with the general expression
compatible with translational and rotational invariance
taken as an initial condition for Eq.~\rf{hck}:
\begin{equation}\label{init}
K_{ij}(0)=\left(\delta _{ij}-\frac{\partial _i\partial
_j}{\partial ^2}\right)K+\frac{\partial _i\partial _j}{\partial ^2}\,K',
\end{equation}
where $K$ and $K'$ are the functions of $\partial ^2$ only. Solving
Eq.~\rf{hck} we find:
\begin{equation}\label{final}
K_{ij}(\xi )=\left(\delta _{ij}-\frac{\partial _i\partial
_j}{\partial ^2}\right)K+\frac{\partial _i\partial _j}{\partial
^2}\,\e^{\xi W\partial ^2}K'.
\end{equation}
Taking
\begin{equation}\label{corw}
\xi W=(\partial ^2)^{-1}(\ln K-\ln K')
\end{equation}
we get the diagonal coefficient function $K_{ij}=\delta _{ij}K$, thus
proving the assertion used in Sec.~\ref{2}.  The extreme case of $\xi
W\rightarrow+\infty$ corresponds to purely transversal coefficient
function, which arises if the functional $S[A]$ is gauge invariant.

\end{document}